# High accuracy, high resolution $^{235}$U(n,f) cross section from n_TOF (CERN) in the thermal to 10 keV energy range.

M. Mastromarco[1], S. Amaducci[2,#], N. Colonna[1], P. Finocchiaro[2], L. Cosentino[2], O. Aberle[6], J. Andrzejewski[7], L. Audouin[8], M. Bacak[9,6,10], J. Balibrea[11], M. Barbagallo[6], F. Bečvář[12], E. Berthoumieux[10], J. Billowes[13], D. Bosnar[14], A. Brown[15], M. Caamaño[16], F. Calviño[17], M. Calviani[6], D. Cano-Ott[11], R. Cardella[6], A. Casanovas[17], F. Cerutti[6], Y. H. Chen[8], E. Chiaveri[6,13,18], G. Cortés[17], M. A. Cortés-Giraldo[18], L. A. Damone[2,19], M. Diakaki[10], C. Domingo-Pardo[20], R. Dressler[21], E. Dupont[10], I. Durán[16], B. Fernández-Domínguez[16], A. Ferrari[6], P. Ferreira[22], V. Furman[23], K. Göbel[24], A. R. García[11], A. Gawlik[7], S. Gilardoni[6], T. Glodariu†[25], I. F. Gonçalves[22], E. González-Romero[11], E. Griesmayer[9], C. Guerrero[18], F. Gunsing[10,6], H. Harada[26], S. Heinitz[21], J. Heyse[27], D. G. Jenkins[15], E. Jericha[9], F. Käppeler[28], Y. Kadi[6], A. Kalamara[29], P. Kavrigin[9], A. Kimura[26], N. Kivel[21], I. Knapova[12], M. Kokkoris[29], M. Krtička[12], D. Kurtulgil[24], E. Leal-Cidoncha[16], C. Lederer[30], H. Leeb[9], J. Lerendegui-Marco[18], S. Lo Meo[3,4], S. J. Lonsdale[30], D. Macina[6], A. Manna[4,5], J. Marganiec[7,31], T. Martínez[11], A. Masi[6], C. Massimi[4,5], P. Mastinu[32], E. A. Maugeri[21], A. Mazzone[2,33], E. Mendoza[11], A. Mengoni[3,4], P. M. Milazzo[34], F. Mingrone[6], A. Musumarra[1,35], A. Negret[25], R. Nolte[31], A. Oprea[25], N. Patronis[36], A. Pavlik[37], J. Perkowski[7], I. Porras[38], J. Praena[38], J. M. Quesada[18], D. Radeck[31], T. Rauscher[39,40], R. Reifarth[24], C. Rubbia[6], J. A. Ryan[13], M. Sabaté-Gilarte[6,18], A. Saxena[41], P. Schillebeeckx[27], D. Schumann[21], P. Sedyshev[23], A. G. Smith[13], N. V. Sosnin[13], A. Stamatopoulos[29], G. Tagliente[2], J. L. Tain[20], A. Tarifeño-Saldivia[17], L. Tassan-Got[8], S. Valenta[12], G. Vannini[4,5], V. Variale[2], P. Vaz[22], A. Ventura[4], V. Vlachoudis[6], R. Vlastou[29], A. Wallner[42], S. Warren[13], C. Weiss[9], P. J. Woods[30], T. Wright[13], P. Žugec[14,6]

[1] Istituto Nazionale di Fisica Nucleare, Sezione di Bari, Italy
[2] INFN Laboratori Nazionali del Sud, Catania, Italy
[3] Agenzia nazionale per le nuove tecnologie (ENEA), Bologna, Italy
[4] Istituto Nazionale di Fisica Nucleare, Sezione di Bologna, Italy
[5] Dipartimento di Fisica e Astronomia, Università di Bologna, Italy
[6] European Organization for Nuclear Research (CERN), Switzerland
[7] University of Lodz, Poland
[8] Institut de Physique Nucléaire, CNRS-IN2P3, Univ. Paris-Sud, Université Paris-Saclay, F-91406 Orsay Cedex, France
[9] Technische Universität Wien, Austria
[10] CEA Irfu, Université Paris-Saclay, F-91191 Gif-sur-Yvette, France
[11] Centro de Investigaciones Energéticas Medioambientales y Tecnológicas (CIEMAT), Spain
[12] Charles University, Prague, Czech Republic
[13] University of Manchester, United Kingdom
[14] Department of Physics, Faculty of Science, University of Zagreb, Zagreb, Croatia
[15] University of York, United Kingdom
[16] University of Santiago de Compostela, Spain
[17] Universitat Politècnica de Catalunya, Spain
[18] Universidad de Sevilla, Spain
[19] Dipartimento di Fisica, Università degli Studi di Bari, Italy
[20] Instituto de Física Corpuscular, CSIC - Universidad de Valencia, Spain
[21] Paul Scherrer Institut (PSI), Villingen, Switzerland
[22] Instituto Superior Técnico, Lisbon, Portugal
[23] Joint Institute for Nuclear Research (JINR), Dubna, Russia
[24] Goethe University Frankfurt, Germany
[25] Horia Hulubei National Institute of Physics and Nuclear Engineering, Romania
[26] Japan Atomic Energy Agency (JAEA), Tokai-mura, Japan
[27] European Commission, Joint Research Centre, Geel, Retieseweg 111, B-2440 Geel, Belgium
[28] Karlsruhe Institute of Technology, Campus North, IKP, 76021 Karlsruhe, Germany
[29] National Technical University of Athens, Greece
[30] School of Physics and Astronomy, University of Edinburgh, United Kingdom
[31] Physikalisch-Technische Bundesanstalt (PTB), Bundesallee 100, 38116 Braunschweig, Germany
[32] Istituto Nazionale di Fisica Nucleare, Sezione di Legnaro, Italy
[33] Consiglio Nazionale delle Ricerche, Bari, Italy
[34] Istituto Nazionale di Fisica Nucleare, Sezione di Trieste, Italy
[35] Dipartimento di Fisica e Astronomia, Università di Catania, Italy
[36] University of Ioannina, Greece

[37]University of Vienna, Faculty of Physics, Vienna, Austria
[38]University of Granada, Spain
[39]Department of Physics, University of Basel, Switzerland
[40]Centre for Astrophysics Research, University of Hertfordshire, United Kingdom
[41]Bhabha Atomic Research Centre (BARC), India
[42]Australian National University, Canberra, Australia

[#] corresponding author, email: amaducci@lns.infn.it

**Abstract**

The $^{235}$U(n,f) cross section was measured in a wide energy range (25 meV – 170 keV) at the n_TOF facility at CERN, relative to $^{6}$Li(n,t) and $^{10}$B(n,α) standard reactions, with high resolution and accuracy, with a setup based on a stack of six samples and six silicon detectors placed in the neutron beam. In this paper we report on the results in the region between thermal and 10 keV neutron energy. A resonance analysis has been performed up to 200 eV, with the code SAMMY. The resulting fission kernels are compared with the ones extracted on the basis of the resonance parameters of the most recent major evaluated data libraries. A comparison of the n_TOF data with the evaluated cross sections is also performed from thermal to 10 keV neutron energy for the energy-averaged cross section in energy groups of suitably chosen width. A good agreement is found in average between the new results and the latest evaluated data files ENDF-B/VIII and JEFF-3.3, as well as with respect to the IAEA reference files. However, some discrepancies are still present in some specific energy regions. The new dataset here presented, characterized by unprecedented resolution and accuracy, can help improving the evaluations in the Resolved Resonance Region and up to 10 keV, and reduce the uncertainties that affect this region.

# 1 Introduction

The neutron-induced fission of $^{235}$U is one of the most important reactions for applications, in particular related to energy production. Its cross section at thermal and from 0.15 to 200 MeV neutron energy is a major standard, widely employed in a variety of fields, from neutron flux measurements to dose evaluation for radiation protection purposes [1]. Outside the standard range, the $^{235}$U(n,f) cross section can also be used as reference, although the presence of resonances and resonance-like structures up to ~10 keV makes the use of this cross section less straightforward. While recently the cross section integral between 7.8 and 11 eV has been adopted as an additional standard, with associated uncertainty close to 1%, uncertainties of the order of a few percent still persist in the Resolved Resonance Region (RRR, corresponding to $E_n$<2.25 keV), as well in the Unresolved Resonance Region (extending up to 25 keV), with some discrepancies between different evaluated data files.

In order to try solving discrepancies in current libraries for $^{235}$U and other key isotopes relevant for nuclear applications, a Collaborative International Evaluation Library Organization (CIELO) was established in 2013, coordinated by the Nuclear Energy Agency (NEA) of the Organization for Economic Cooperation and Development (OECD). A detailed description of the project and related results can be found in [2]. Although progresses were made within the project for several reactions, open questions and differences in the evaluations still remain, documented in two different datasets, CIELO-1 and CIELO-2, adopted by different evaluated data libraries. Differences of the order of a few percent still persist on some crucial reactions, including the $^{235}$U(n,f) reaction outside the standard region. In an attempt to reduce the uncertainties, new collaborative efforts are being undertaken, such as the INDEN project coordinated by IAEA [3], that aims at improving the evaluation methodology and producing updated nuclear data files. In this respect, while re-analysis and combination of previous data can lead to some improvements, a major uncertainty reduction can be achieved by incorporating new, high resolution and high accuracy data, that can help sorting-out existing discrepancies. In this respect, the n_TOF facility [4] is currently one of the best suited facilities worldwide for collecting new data on the $^{235}$U(n,f) cross sections in the Resolved and Unresolved Resonance Regions, thanks to the very convenient features of the neutron beam, in particular the high resolution, the wide energy range and the low background. An overview of the facility and of the fission experimental program at n_TOF can be found in [5].

Data from n_TOF in the RRR collected with Parallel Plate Avalanche Counters (PPAC) [6], were already made available several years ago, and have been used in recent evaluations. In average they showed a good agreement up to 2 keV with the IAEA reference file (being ~1% lower), while a larger difference was observed relative to ENDF/B-VII (being ~2% higher). A difference of 3% has also been reported by Capote et al. in Ref. [7] between ENDF/B-VII and the new IAEA CIELO evaluated cross section between 100 eV and 2.25 keV (the latter being consistent with the IAEA 2017 reference file, see Figure 2 in [7]). It should be noted that the new ENDF/B-VIII evaluation, officially released in 2018 [8] [9], has adopted the IAEA CIELO evaluations, so that the two files are now consistent with each other for the considered reaction. The n_TOF PPAC data in the RRR [6] [10] have also been used in the new JEFF-3.3 evaluation, officially released in 2017 [11] [12]. However, as will be shown in this work, albeit small, differences between ENDF/B-VIII and JEFF-3.3 on the $^{235}$U(n,f) cross section up to 10 keV still persist, reflecting the differences between the CIELO-1 and CIELO-2 evaluated files.

The $^{235}$U(n,f) n_TOF PPAC data mentioned above were collected in the long flight-base experimental area (EAR1), and are therefore characterized by a high resolution ($\Delta E/E<10^{-3}$). However, those data were not the result of a dedicated measurement relative to a standard, such as the $^{6}$Li(n,t) or $^{10}$B(n,$\alpha$), being the $^{235}$U sample used as reference for other actinide samples measured simultaneously. Rather, the cross section was extracted relative to the neutron flux that had previously been determined on the basis of various standards, including the $^{235}$U(n,f) reaction itself, with a relatively low resolution and an uncertainty of up to 5% in the keV region [13]. Furthermore, the energy range did not extend down to thermal energy, hindering an accurate normalization to a well established standard. As a consequence, the data were normalized to the cross section integral in the region between 7.8 and 11.0 eV, and the value of 246.4 b·eV in the IAEA reference file was used to that purpose (at that time not yet adopted as an additional standard, the currently adopted standard value being 247.5 b·eV). As mentioned in Ref. [6], the discrepancy between the n_TOF normalized PPAC data and ENDF/B-VII was indeed mostly related to a 2% difference in the value of the integral cross section used for normalization between that library and the IAEA reference file.

Recently, a measurement was performed at n_TOF specifically dedicated to the high-resolution, high-accuracy measurement of the $^{235}$U(n,f) reaction in the whole energy region from thermal to 170 keV neutron energy. The main aim of that measurement was to investigate a discrepancy that had previously been noted in the 10-30 keV energy range. Details on the measurement, the experimental setup and the analysis procedure can be found in Ref. [14], where the average cross section is reported in the 10-30 keV range and the results discussed. The main features of this new dataset is that the $^{235}$U(n,f) cross section is measured directly relative to the $^{6}$Li and $^{10}$B standards, and that the energy range encompasses the thermal point and extends slightly above 150 keV, i.e. in the standard regions, so that a high confidence on the absolute normalization can be achieved. The present paper complements the previous publication [14], by reporting the pointwise data in the range from thermal to 10 keV, in an attempt to provide some of the most accurate and high-resolution data achieved so far on the $^{235}$U(n,f) cross section in the Resolved and Unresolved Resonance Region. The results here reported could help solving existing discrepancies between different evaluations, providing an important contribution to future upgrades of evaluated libraries and/or ongoing collaborative efforts, such as the INDEN project of IAEA, with the final goal of improving the accuracy of this important cross section in the mentioned energy region.

The paper is organized as follows: the main features of the experimental setup and procedure are described in Section 2. Results are presented in Section 3. A resonance analysis up to 200 eV is reported in 3.1, together with a comparison of the resulting fission kernels with current versions of major libraries. Energy-averaged cross section data from thermal to 10 keV neutron energy are presented and discussed in Section 3.2, in comparison with current and past evaluations and previous experimental data. Conclusions are drawn in Section 4.

## 2. Experimental setup and data analysis

The experimental apparatus consists of a stack of six Si-detectors with six samples, two of each $^{6}$Li, $^{10}$B and $^{235}$U isotope, mounted in between in a closely-packed geometry [14]. The main features of the setup were: a large solid angle coverage, the possibility to detect reaction products both in the forward and backward direction, a feature particularly important for the $^{6}$Li(n,t) and $^{10}$B(n,$\alpha$) reactions, affected by angular anisotropy of the emitted charged particle that starts becoming relevant for neutron energy above a few keV. Finally, the setup was characterized by a good particle identification capability, of importance for background rejection. The setup was hosted in a vacuum chamber placed on the neutron beam line, with thin windows at the air-vacuum interface. The measurement was performed in the experimental area (EAR1) positioned at the end of the long neutron flight path ($\approx$ 184 m). In this area, the energy resolution of the neutron beam is of the order of $10^{-4}$ up to a few keV.

The $^{235}$U(n,f) cross section was determined directly relative to the $^{6}$Li(n,t) and $^{10}$B(n,$\alpha$) reactions, whose cross sections are standards of measurement all the way from thermal to 1 MeV. Another important feature of the measurement was the large energy range covered, extending from thermal neutron energy to 170 keV. Relative to the n_TOF PPAC data mentioned above (already available on EXFOR), the new measurement has the advantage of including both the thermal point and an energy region above 150 keV, where the $^{235}$U(n,f) cross section is standard, thus allowing for an accurate absolute normalization of the data in the whole energy range measured. In fact, the dataset here presented is one of the most complete in terms of energy range covered, and accurate in terms of reference reactions used, as well as for absolute normalization. Combined with the high resolution and low background of the n_TOF neutron beam in EAR1, the wide energy range and high accuracy makes the present dataset rather unique in the landscape of experimental cross sections available on the $^{235}$U(n,f) reaction in the Resolved and Unresolved Resonance Regions, and suitable for verifying current major evaluations, possibly identifying residual problems (if any). To this aim, we perform here a thorough comparison of the data with three libraries: ENDF/B-VII, ENDF/B-VIII and JEFF-3.3 (evaluated cross sections in the JENDL-5 library are similar to ENDF/B-VIII.0 and are therefore not included in this comparison).

## 3. Results

The yield with coarse energy binning from thermal to 170 keV neutron energy has already been reported in Ref. [14] and uploaded on EXFOR. In this paper we report on the analysis of the data, up to 10 keV, with the higher resolution required to perform a resonance analysis and a more detailed comparison with evaluated data libraries. These data will then be made available on EXFOR for dissemination and possible use in future evaluations of the $^{235}$U(n,f) reaction in the Resolved and Unresolved Resonance Regions.

The main features of the present data, in terms of background, absolute normalization and uncertainty estimates, have mostly been discussed in [14]. We recall here that, thanks to the characteristics of the n_TOF facility and the low mass of the detectors, the measurement is affected by a negligible background. This feature is particularly important in the valleys between resonances, as it leads on the one hand to a more reliable resonance analysis in the tails, and on the other hand to a more accurate determination of the energy-averaged cross section, both affected by a non-negligible contribution of the valleys. The low background of the measurement is evident in Figure 1, where the n_TOF data in the valleys are comparable or lower than the evaluated cross sections, showing in some regions structures not present in the evaluations, and that could have been previously masked by a higher background.

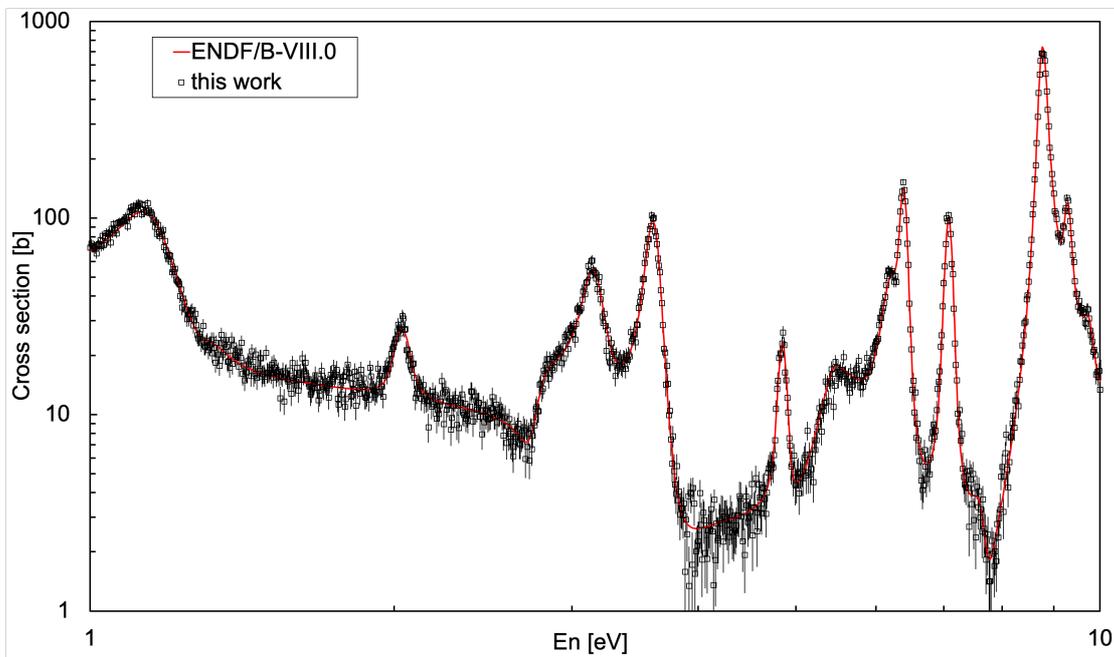

Figure 1: Measured $^{235}$U(n,f) cross section in the 1-10 eV neutron energy range, compared with the ENDF/B-VIII evaluated data. The high resolution of the n_TOF data makes possible to distinctly observe small structures in all the valleys between resonances, clearly testifying the low background and possibly indicating the presence of interference patterns.

As already mentioned, an important feature of the present data is that the cross section is extracted directly from the ratio of the measured count-rate for the $^{235}$U sample to the reference $^6$Li and $^{10}$B samples. This procedure completely removes the effect of structures, in particular absorption dips, that are typically present in the neutron flux, related to the neutron source itself or to windows at the air-vacuum interface on the neutron beam line. At n_TOF, the presence of a thick Al window produces several larger dips, while a 0.2% Zn content produces smaller dips that however could lead to a few percent error in the cross section determination (as will be shown later).

Finally, a fundamental feature of the present data (compared to most previous datasets) is that, thanks to the wide energy range covered in a single measurement, it is possible to verify the absolute normalization of the data against the $^{235}$U(n,f) cross section standard in two regions, i.e. at thermal and around 150 keV, thus providing high confidence on the cross section values here reported. As discussed in [14], the absolute normalization of the measured cross section was performed relative to the cross section integral in the neutron energy range between 7.8 and 11.0 eV, recently adopted as a standard (247.5 b·eV) [15]. An overall uncertainty close to 1.5% has been estimated for the whole range here reported, making the present measurement one of the most accurate ever performed on this reaction.

In the following, we present the results and a comparison with evaluated nuclear data of major libraries, by performing on the one hand a resonance analysis at lower energy, and averaging the cross section in statistically significant energy binning in a wider energy range. The new, high accuracy and high resolution data on this important reaction could be beneficial for checking the reliability of current evaluations in major data libraries, possibly identifying residual shortcomings for future updates.

## 3.1 Resonance analysis

A resonance analysis was performed with the SAMMY code [16], within the Reich-Moore approximation, from thermal to 200 eV. Although the limit of the Resolved Resonance Region is currently assumed at 2.25 keV, and resonance structures are observed even above this limit, in the present work we have considered only resonances up to 200 eV, since at higher energy the clustering of resonances becomes dominant, and the number of missing levels increases. Furthermore, up to this energy the n_TOF neutron beam resolution function has a negligible influence, being the resolution essentially dominated by Doppler broadening. It should be considered that even at this low energy, several resonance structures are made of clusters of unresolved resonances, due to the small average level spacing, as compared to their natural width and to the effect of the Doppler broadening. Finally, above 200 eV a more statistically meaningful comparison with libraries and previous datasets can be performed by averaging the cross section in suitably wide energy bins as discussed later on.

In the SAMMY fits, initial input parameters of the resonances were taken from the latest evaluated data files, either ENDF/B-VIII or JEFF-3.3. The energy, neutron and capture widths were kept fixed, while both fission widths were left free, with a fudge factor set at 0.1 (corresponding to the possibility to modify the value by 10% for each iteration). The numerical n_TOF resolution function was used in the fits. In order to determine the normalization value of the n_TOF yields, the corresponding parameter was left free when fitting the energy region from 20 meV to 10 eV. The values obtained using initial resonance parameters from the ENDF/B-VIII and JEFF-3.3 where 0.993 and 0.997, respectively, indicating that the absolute normalization of the n_TOF yield was consistent with evaluated cross sections within a few per thousand. For the fits from 10 to 200 eV neutron energy, the normalization parameters were kept fixed at the values mentioned above. For the resonance fits in the whole neutron energy region analysed, the level of background was left free, but the resulting values were in all cases negligible. Figure 2 shows the experimental data together with the results of the SAMMY fit in the whole range from thermal to 200 eV neutron energy.

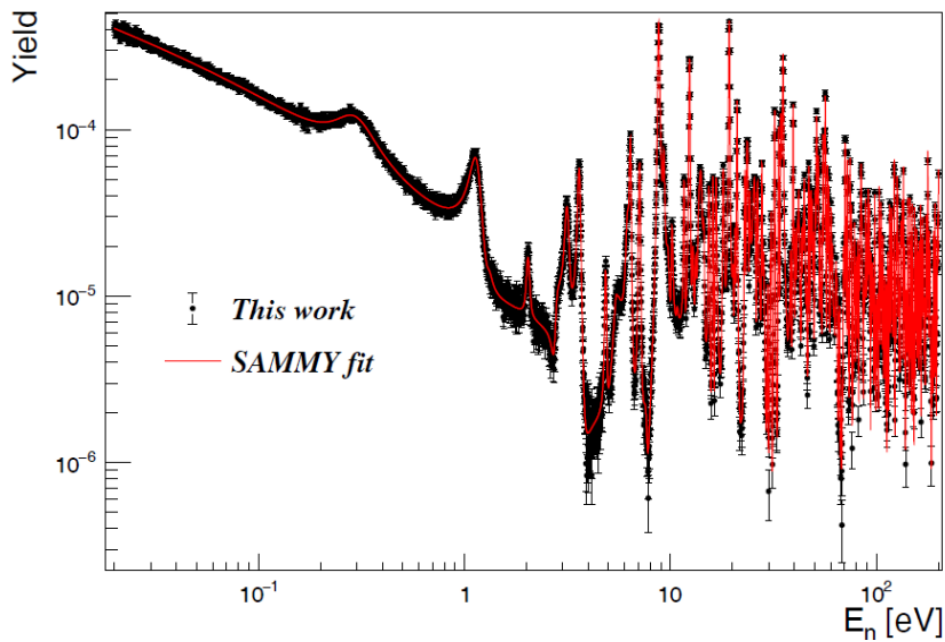

Figure 2: Yields of the $^{235}$U(n,f) reaction measured at n_TOF (symbols), together with the result of resonance analysis performed with the SAMMY code up to 200 eV.

The quality of the fits in selected energy regions can be appreciated in Figure 3. The experimental fission yield and the result of the resonance analysis performed within this work are represented by the symbols, with their statistical uncertainty, and by the red curve, respectively. They are compared with the fission yield calculated by SAMMY on the basis of the resonance parameters of the two most recent evaluated data libraries, represented in the figures by the green curve (for clarity, the comparison with ENDF/B-VIII and JEFF-3.3 is shown separately). For most resonances, the new fit and evaluated fission yields are indistinguishable, indicating a very good agreement between the n_TOF data reported here and the evaluated cross sections. In some cases, however, the evaluations fall short of reproducing the observed resonances, as can be inferred from the subtle differences between the red and the green curve in Figure 3. Most of the differences are observed in the neutron energy range between 20 to 100 eV (as confirmed by the analysis of the energy-averaged cross section, shown later). The ranges in Figure 3 correspond to the region where larger differences between data and evaluations are present.

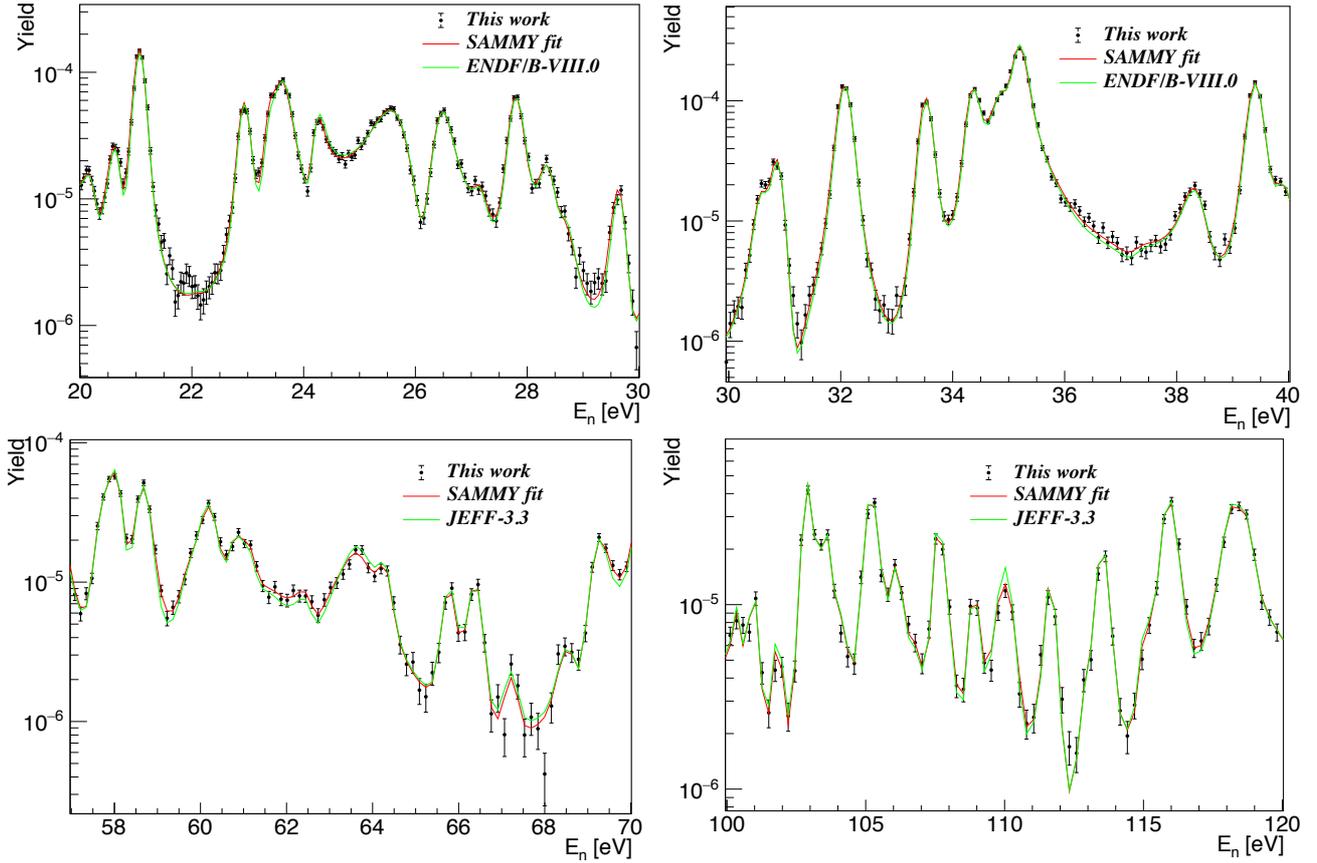

Figure 3: Example of resonance fits of $^{235}$U(n,f) resonances from the present n_TOF yield data (symbols and green curves), compared with yields calculated in SAMMY from resonance parameters of the two most recent libraries, ENDF/B-VIII and JEFF-3.3. Where differences exist, the red and green curve depart from each other.

A more quantitative comparison can be performed by considering the fission widths. However, although this is clearly the most important parameter to be determined from the resonance analysis of fission data, the values of such width are correlated to the neutron and, to a lesser extent, the capture widths, which may be different in the two major libraries. As a consequence, it is more appropriate, when comparing the experimental data with both libraries, to consider the fission kernel defined as follows,

$$K_f = g \, \frac{\Gamma_n(\Gamma_{f1} + \Gamma_{f2})}{\Gamma_n + \Gamma_\gamma + \Gamma_{f1} + \Gamma_{f2}}$$

Here $g$ is the spin factor, $\Gamma_n$ and $\Gamma_\gamma$ are the neutron and capture width, respectively, while $\Gamma_{f1}$ and $\Gamma_{f2}$ the two fission widths. Figure 4 shows the ratio of the experimental fission kernels to the ones extracted using the ENDF/B-VIII and JEFF-3.3 resonance parameters. The error bars represent the uncertainty in the SAMMY fit, and are essentially related to the statistical uncertainty on the n_TOF fission yield data. A very good agreement is observed in average between the present data and both major libraries. Indeed, a weighted average of 1.0037(3) and 1.0016(2) is found for the ratio relative to ENDF/B-VIII and JEFF-3.3, respectively. However, as already mentioned, differences of several percent can be observed for some resonances (or cluster of resonances), mostly small ones, as can be inferred from the large error bars. Nevertheless, the weighted root mean square of the ratio, that provides an indication of the width of the ratio distribution, turns out to be also small, being less than 1% for both libraries (0.0096 and 0,0064 for ENDF/B-VIII and JEFF-3.3, respectively). This further implies that the two libraries are essentially equivalent and in average closely reproduce the observed n_TOF resonances up to 200 eV.

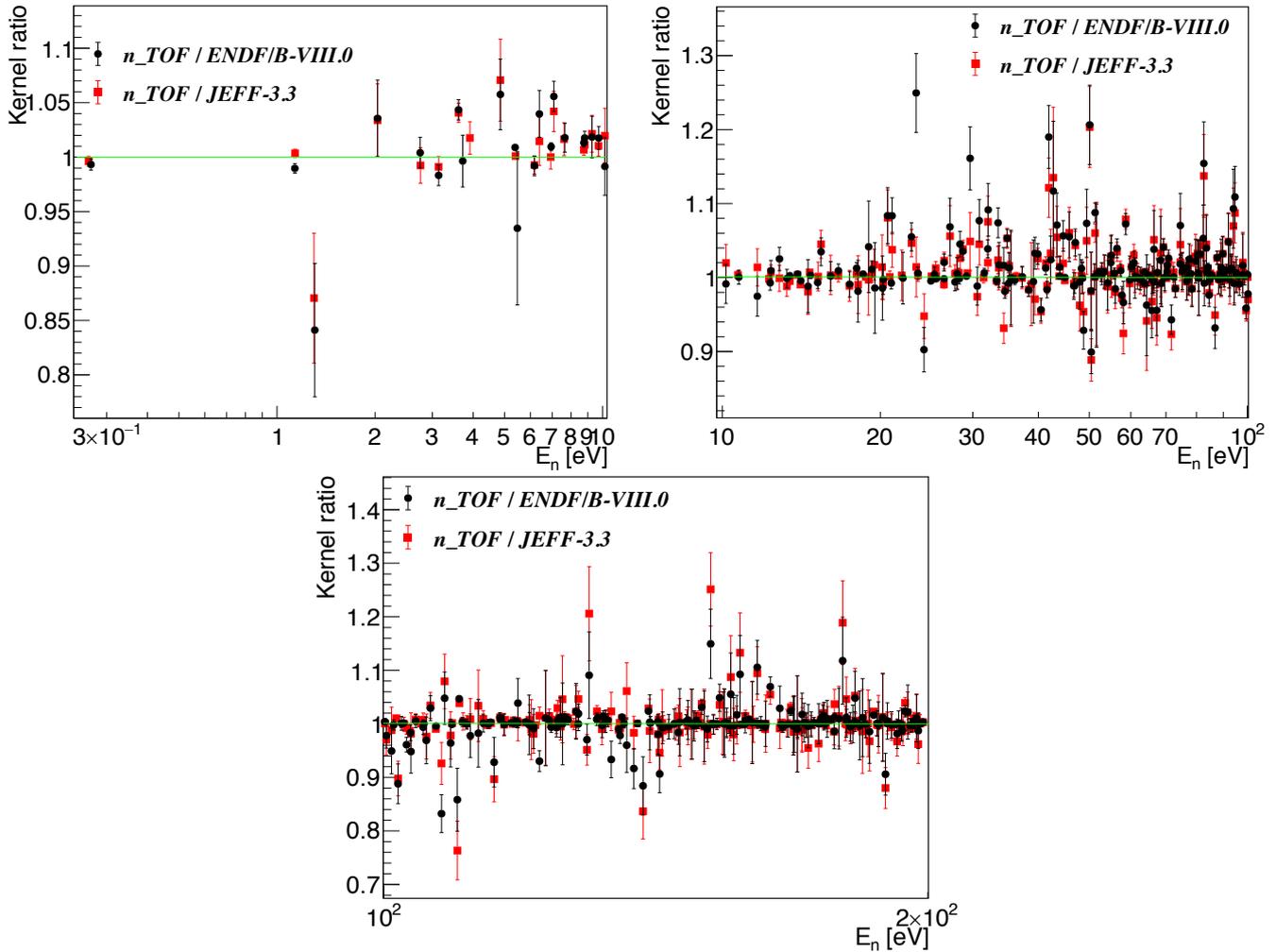

Figure 4: Ratio of the n_TOF fission kernels (determined from SAMMY fit of the resonances), and those calculated from the resonance parameters of the two most recent evaluated data libraries (ENDF/B-VIII and JEFF-3.3). See text for details.

**3.2 Energy-averaged fission cross sections from thermal to 10 keV neutron energy**

While the resonance analysis shown above can provide some indications on the accuracy of the evaluations up to 200 eV, a more complete comparison can be performed, all the way from thermal to 10 keV neutron energy, by considering the average cross sections, in energy bins (or groups) of suitably chosen width. Comparison in this sense have been reported in the literature both between data and libraries (see for example Ref. [6]) and between evaluated data files [7] [11]. In particular, in Refs. [6] and [7], it was shown that the cross sections in ENDF/B-VII were higher than the IAEA reference file and the n_TOF PPAC data, by several percent. Following the CIELO project, a modification was introduced in the ENDF/B-VIII, which adopted the CIELO-1 evaluations (mostly based for this reaction on the IAEA reference file). Similarly, the new version of JEFF-3.3 was shown to reproduce the latest standards evaluations [1], as well as the previous one [17] in the neutron energy range 100-2000 eV (see Table 3 in [11]).

Taking advantage of the wide energy range of the present n_TOF data, we have performed a complete comparison with the two current evaluations, as well as with the previous ENDF/B-VII, with the aim of verifying improvements and/or residual differences. The comparison over the full energy range measured at n_TOF is shown in Figure 5, where the reported R values represent the discrepancy between the current data and the evaluations. For a more detailed analysis the whole neutron energy region from thermal to 10 keV has been divided in different panels in Figure 6. It has to be considered, however, that between 10 and 30 keV the new evaluations show still an important discrepancy relative to the n_TOF data, as already discussed in Ref. [14]). The size of the energy bin has been chosen for all cases to be 10 bins/decade, a value that fulfils the best compromise between the need of a reasonable resolution and that of minimizing statistical fluctuations, in particular in the keV energy region.

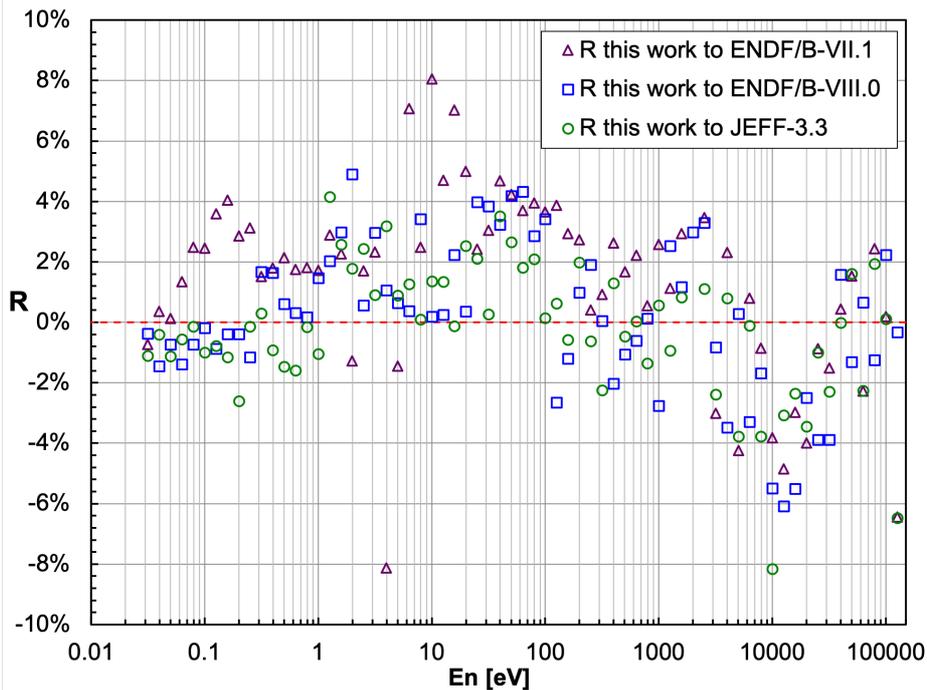

Figure 5: Ratio of the $^{235}$U(n,f) energy-averaged cross section measured at n_TOF in the full neutron energy range, from thermal to 170 keV, and the evaluated cross section from major current data libraries, namely ENDF/B-VIII and JEFF-3.3. The ratio to the older ENDF/B-VII.1 library is also shown for comparison. The average was calculated over 10 bins/decade.

The measured energy-averaged cross section was obtained from the fission yield dividing by the number of atoms/barns. In this respect, the self-absorption correction is not taken into account, as the small thickness of the $^{235}$U samples used in the n_TOF measurement results in a negligible effect, of the order of $10^{-4}$. For the evaluations, the energy-averaged cross section was derived from the pointwise values, by first interpolating them (according to the library prescriptions) at a very fine resolution (2000 bins/decade), and then integrating the resulting cross section over the chosen energy intervals corresponding to the experimental data (i.e 10 bins/decade). In the plots, a generally good agreement of the new libraries with the n_TOF data can be observed, in the whole energy region from thermal to 10 keV. On the contrary, the older ENDF/B-VII library shows a rather large systematic discrepancy, being the n_TOF data considerably higher than the evaluations by several percent, from a few hundred meV to a few keV. While the underestimate of the cross section in ENDF/B-VII in the recently established standard from 7.8 to 11.0 eV had been previously reported, as compared to the IAEA reference value (see for example Ref. [6]), a somewhat unexpected finding is the discrepancy in the 100-300 meV neutron energy. In these two energy ranges, the recent re-evaluation within the CIELO project, adopted in the most recent libraries, has led to a substantial improvement of the cross section, relative to previous evaluated data, as demonstrated by the very good agreement with the present n_TOF data. However, some discrepancies can still be noted even for the new libraries. At low energy, up to approximately 1 eV, the n_TOF cross sections are slightly lower than JEFF-3.3 evaluations, by around 1%, while they are in good agreement with the ENDF/B-VIII evaluations. On the contrary, the two libraries mostly agree in the range 1-100 eV, but the n_TOF data are systematically higher than both of them at a few eV and between 20 and 80 eV. It is interesting to note that in that range there has been no change in the new evaluations relative to the previous ones (see comparison with ENDF/B-VII). It should be also noted that, as shown previously, important differences are observed in the resonances in that range, therefore calling for a better evaluation of the resonance parameters in future library releases.

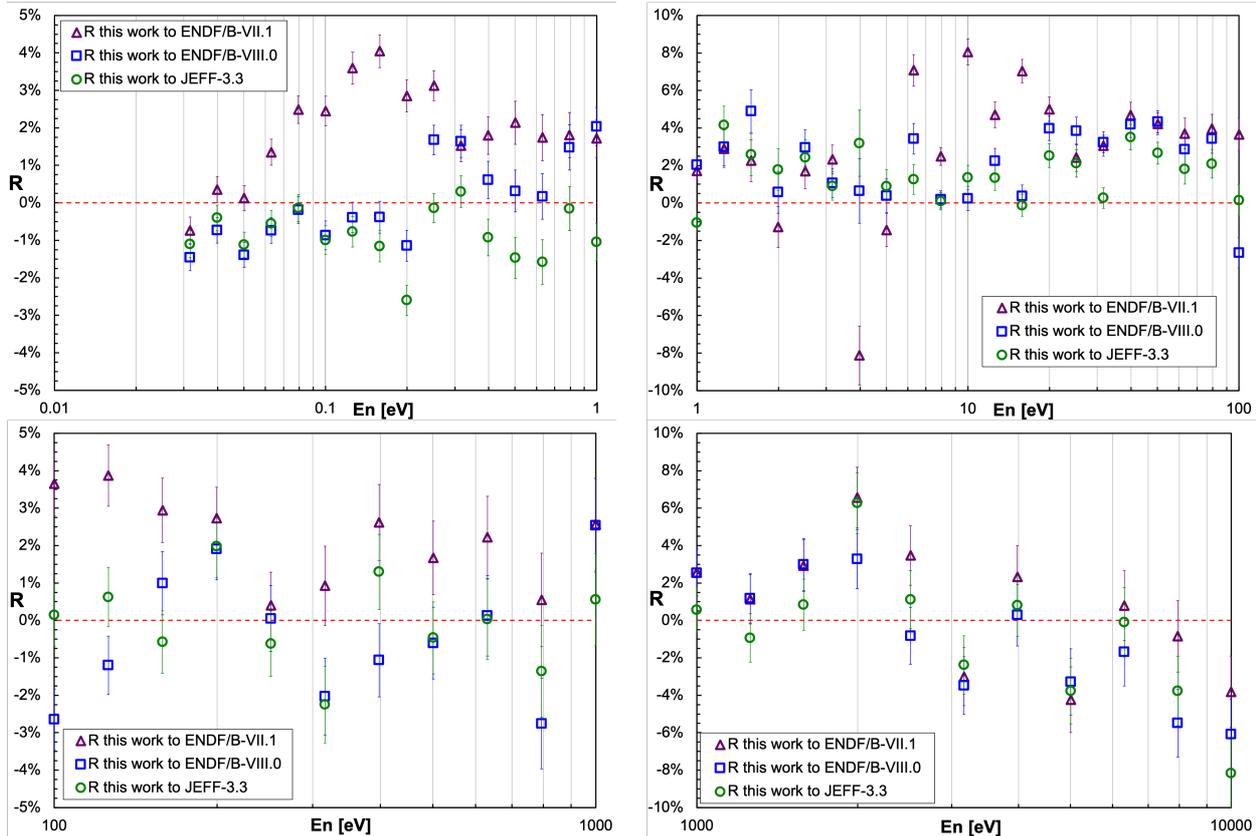

Figure 6: Comparison of the present n_TOF cross section averaged in 10 bins/decade, with energy-averaged cross sections from the two most recent evaluated data libraries, ENDF/B-VIII and JEFF-3.3. The older ENDF/B-VII library is also included in the comparison, showing the substantial improvement in current evaluations relative to previous ones. In some energy regions, difference of several percent are still present in current evaluated files. The four panels represent different neutron energy ranges, from thermal to 1 eV (upper left panel), from 1 to 100 eV (upper right), from 100 eV to 1 keV (bottom left) and finally from 1 to 10 keV (bottom right). The purple triangles show the ratio between the present data and ENDF/B-VII, the blue squares the ratio to ENDF/B-VIII and the green circles the one to JEFF-3.3.

In the 100 eV to 1 keV range, no particular problems are observed, except for a 2% discrepancy around 200 eV neutron energy. Finally, the comparison in the 1 to 10 keV range signals a further problem around 2 keV. This energy corresponds approximately to the limit of the Resolved Resonance Region in the libraries, and at least part of the observed discrepancy could be attributed to the difference of treatment in the evaluation between the RRR and URR. After a careful investigation, however, we have come to the conclusion that most probably this is not the reason for the differences between the present n_TOF data and evaluations. A comparison with previous data, in particular the n_TOF PPAC data of Ref. [6], shows a large difference between 2 and 3 keV (a difference is also observed, relative to Weston and Todd [18], although smaller). In this region, the n_TOF flux exhibits a small dip, as it can be appreciated in the neutron flux determined in the measurement here reported from the $^6$Li(n,t) and $^{10}$B(n,$\alpha$) reactions (see Figure 11 of Ref. [14]). The dip can be attributed to the 2.63 keV resonance in the total cross section of $^{64}$Zn (dominated by elastic scattering). Zinc is present in the neutron source at n_TOF, being a 0.2% contaminant of the Al alloy 6082, used for the window at the interface between the spallation target and the vacuum beam line. Being the dip small and barely visible in the adopted energy distribution of the neutron beam in EAR1 [13], it was not accounted for in the analysis of Ref. [6], in which the cross section was extracted relative to that energy distribution. It can therefore be concluded that the observed difference relative to the IAEA reference file observed in Ref. [6] around 2 keV neutron energy is essentially related to the mentioned absorption dip, and it cannot be excluded that a similar problem might also be present in other previous datasets, as already mentioned. As it can be noted in Figure 6, in the recent evaluations the cross section have been increased in that energy bin, and agree better with the present data. A reminiscence of the problem might however still be present at slightly lower energy, being the n_TOF data in the bin at 2 keV higher, by 3% and 6% relative to ENDF/B-VIII and JEFF-3.3, respectively. A re-evaluation of the cross section in this energy region might therefore be needed to account for the new evidences here discussed.

The overall performance of the two most recent evaluated data libraries, in reproducing the new n_TOF data reported here can be inferred from Figure 7, that shows the weighted average deviation of the evaluated cross sections from the present data, in the full energy range examined, i.e. between thermal and 10 keV, and in four sub-ranges. From the figure, one can appreciate the large improvements in the most recent libraries, following the CIELO project, relative to the previous one. Overall, both ENDF/B-VIII and JEFF-3.3 agree with the new data within less than half a percent, a remarkable result, in particular if compared with the systematic difference of the present data relative to ENDF/B-VII by 2%. However, it should be noted that such a good overall agreement with the most recent evaluations is the result of positive and negative deviations in different energy regions that compensate each other. Nevertheless, in all ranges the agreement with the two current evaluated data libraries is around 1%, except for the 1-100 eV range, where a 2% difference is observed for ENDF/B-VIII, mostly related to a problem in the 20-80 eV region, as also shown in the comparison of the resonance kernels and in Figure 6.

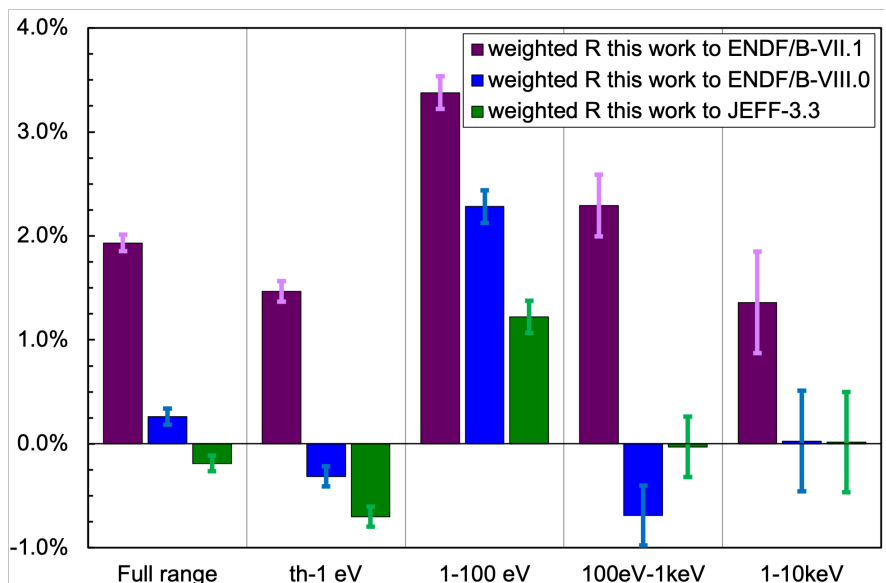

Figure 7: Weighted average ratio of the n_TOF data of this work to the most recent evaluations, as well as a previous version of the ENDF library, in the overall energy range from thermal to 10 keV, and in smaller ranges.

It is also interesting to compare the present data with a selection of previous experimental datasets, that have been considered in past and current evaluations. In particular, the comparison has been done relative to the n_TOF PPAC data of Paradela et al. [6], and the earlier measurements of Weston [18] and Mostovaya et al. [19]. Figure 8 shows the ratio of the n_TOF data of this work to previous results. It should be considered that the range covered by the present and earlier datasets do not completely overlap, being 0.66 eV to 10 keV for Paradela et al. [6], 9.73 eV to 200 keV for Weston and Todd [18] and 78 eV to 20 keV for Mostovaya et al. [19]. In this case, the comparison was performed only in the region where the previous data overlap, totally or partially, with the present ones. All previous dataset considered in this work have no data from thermal to 1 eV energy, so that no comparison could be performed in that region. Relative to the data of Weston and Todd [18], the n_TOF results presented here are systematically higher by 1 to 2%, while they are higher relative to the previous n_TOF PPAC data [6] by 0.7% in the 1-100 eV range, and up to 1.8% in the 1-10 keV region. This difference is most probably related on the one hand to a slightly less accurate normalization of the n_TOF PPAC data, and on the other hand to the larger uncertainty in the neutron flux used as reference for extracting the cross section, in particular above 1 keV, in the n_TOF PPAC dataset. Finally, an almost perfect overall agreement, within 1%, can be observed between present data and the ones from Mostovaya et al. [19], although this is the result of a combination of a large positive difference from 1 to 100 eV, and the smaller negative one at higher energy (of a few per thousand from 100 eV to 1 keV and around 1% in 1-10 keV range).

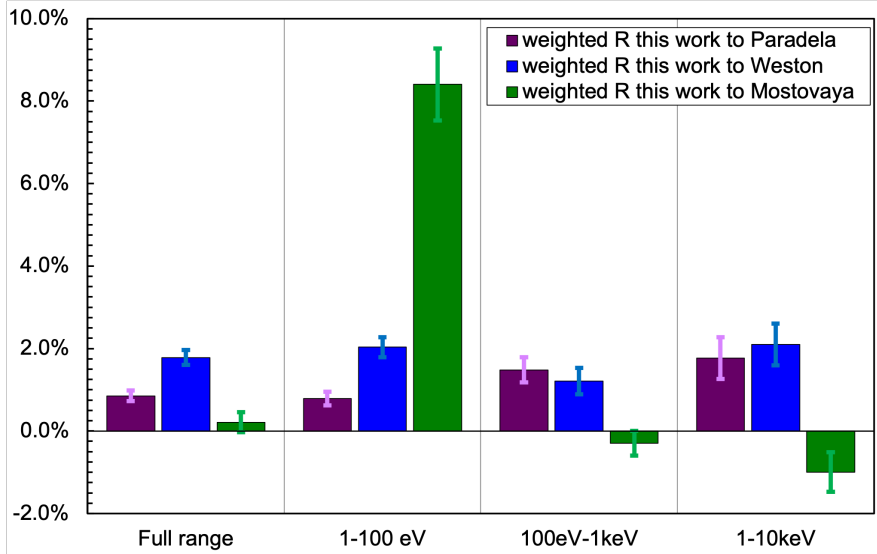

Figure 8: Weighted average ratio between the present dataset and a selection of previous experimental results: previous n_TOF PPAC data of Paradela et al., [6], Weston and Todd [18] and Mostovaya et al., [19]. In the range from thermal to 1 eV neutron energy it is not possible to perform a comparison in some cases, because of the lack of data in that region for earlier measurements.

The n_TOF energy-averaged cross sections from 100 eV to 2 keV in 100 eV wide energy groups are compared with the IAEA reference files and with the JEFF-3.3 evaluations in Table 1 (that essentially reproduces Table 3 of Ref. [11] with the addition of the present results). The table also includes the energy-integrated cross section between 7.8 and 11 eV, now adopted as an additional standard. It is interesting to note the almost perfect agreement between the present data and the IAEA reference files of 2009 and 2018, that have now been adopted in the latest version of the ENDF/B library, as well as with the new JEFF-3.3 evaluation, demonstrating the reliability in average of this last library as well as of the IAEA reference file.

Table 1: Energy-averaged cross section from 100 eV to 2 keV, calculated from the IAEA standards of 2009 [17] and 2018 [1], the latest JEFF-3.3 evalutions (all from Table 3 of Ref. [11]), and the n_TOF data reported in this work. The integrated cross section in the 7.8-11.0 eV range, used in this work for absolute normalization, is also included in the Table.

|  | IAEA 2009 | IAEA 2018 | JEFF-3.3 | this work 1000 bpd |
|---|---|---|---|---|
| (eV) | (b eV) | (b eV) | (b eV) | (b eV) |
| 7.8–11 | 246.4(12) | 247.5(30) | 246.9 | 247.5(1.1) |
| (eV) | (b) | (b) | (b) | (b) |
| 100–200 | 21.17(11) | 21.3(3) | 21.02 | 21.07(10) |
| 200–300 | 20.69(11) | 20.8(3) | 20.77 | 20.99(14) |
| 300–400 | 13.13(7) | 13.2(2) | 13.22 | 13.04(13) |
| 400–500 | 13.78(8) | 13.9(2) | 13.49 | 13.66(14) |
| 500–600 | 15.17(9) | 15.2(2) | 15.2 | 15.16(17) |
| 600–700 | 11.51(7) | 11.57(15) | 11.53 | 11.55(15) |
| 700–800 | 11.10(6) | 11.15(14) | 11.1 | 11.19(16) |
| 800–900 | 8.21(5) | 8.25(11) | 8.15 | 8.01(14) |
| 900–1000 | 7.50(4) | 7.54(10) | 7.37 | 7.42(14) |
| 1000–2000 | 7.30(4) | 7.34(10) | 7.29 | 7.36(06) |

## 4. Conclusions

The $^{235}$U(n,f) cross section was determined with high accuracy and high resolution from thermal to 170 keV, in a dedicated measurement performed at n_TOF in the experimental area at the end of the 184 m long flight-path. The cross section was extracted from a direct ratio measurement, relative to the the $^6$Li(n,t) and the $^{10}$B(n,$\alpha$) standard reactions. A compact, Si-based setup placed in the neutron beam was used for the detection of the reaction products. Data have been collected both in the forward and in the backward direction, to minimize the uncertainty related to angular anisotropy in the charged particle emission from the reference reactions. An absolute normalization has been performed for all three reactions in the 7.8-11.0 eV energy region, using the recommended IAEA standard. The resulting cross sections were checked against the standard at thermal and 150 keV neutron energy. This procedure has resulted in an unprecedented low uncertainty of 1.5%. Such a high accuracy, combined with the high resolution and the wide energy range covered, makes the data reported here among the most complete and reliable ever collected in this RRR and URR.

The results have been compared with recent evaluated data libraries, that have included the outcome of the CIELO project, devoted to improving cross section evaluations for applications. In particular, we have compared resonance fission kernels and energy-averaged cross sections with the values extracted from the two most recent evaluated libraries, ENDF/B-VIII and JEFF-3.3, as well as with the previous ENDF/B-VII library. An overall good agreement is observed, within 1%, with the new evaluations, with a significant improvement relative to previous versions. However, a more detailed comparison shows that discrepancies still exist between present data and all evalutions in some specific energy regions, namely at a few eV, between 20 and 80 eV, around 2 keV. A previously indicated discrepancy at 10-30 keV is also still present. The availability of the present data on EXFOR might help improving future evaluations in these regions, solving remaining discrepancies.

**Acknowledgments**

**References**

[1] A.D. Carlson et al., *Nucl. Data Sheets,* vol. 148, p. 143, (2018).
[2] M.B. Chadwick et al., *Nuclear Data Sheets,* vol. 148, p. 189, (2018).
[3] [Online]. Available: https://www-nds.iaea.org/INDEN/.
[4] C. Guerrero et al., *Eur. Phys. J. A,* vol. 49, p. 27, (2013).
[5] N. Colonna et al., *Eur. Phys. J. A,* vol. 56, p. 48, (2020).
[6] C. Paradela et al., *EPJ Web of Conferences,* vol. 111, p. 02003, (2016).
[7] R. Capote et al., *Nucl. Data Sheets,* vol. 148, p. 254, (2018).
[8] D. A. Brown et al., *Nuclear Data Sheets,* vol. 148, p. 1, (2018).
[9] [Online]. Available: https://www.nndc.bnl.gov/endf-b8.0/.
[10] I. Duran et al., *EPJ Web of Conferences,* vol. 211, p. 02003, (2019).
[11] A. Plompen et al., *Eur. Phys. J. A,* vol. 56, p. 181, (2020).
[12] [Online]. Available: https://www.oecd-nea.org/dbdata/jeff/jeff33/.
[13] M. Barbagallo et al., *Eur. Phys. J. A,* vol. 49, p. 156, (2013).
[14] S. Amaducci et al., *Eur. Phys. J. A,* vol. 55, p. 120, (2019).
[15] [Online]. Available: https://www-nds.iaea.org/standards/.
[16] N. M. Larson, *Report ORNL/TM-9179/R8,* (2008).
[17] A. D. Carlson et al., *Nucl. Data Sheets,* vol. 110, p. 3215, (2009).
[18] J. T. L.W. Weston, *Nucl. Sci. Eng. ,* vol. 88, p. 567, (1984).
[19] T.A. Mostovaya et al., *Proceedings of the 5th All Union Conference on Neutron Physics,* vol. 3, pp. 30-32, (1980).